\documentclass{vgtc}  

\vgtccategory{Research}

\title{ViStruct: Simulating Expert-Like Reasoning Through Task Decomposition and Visual Attention}

\author{\authororcid{Oliver Huang}{0009-0007-1585-1229}\thanks{e-mail: oliver@cs.toronto.edu}\\ %
        \scriptsize University of Toronto %
\and \authororcid{Carolina Nobre}{0000-0002-2892-0509}\thanks{e-mail: cnobre@cs.toronto.edu}\\ %
     \scriptsize University of Toronto %
}

\abstract{%
  Data visualization tasks often require multi-step reasoning, and the interpretive strategies experts use—such as decomposing complex goals into smaller subtasks and selectively attending to key chart regions—are rarely made explicit. We developed ViStruct as an automated pipeline that simulates these expert behaviours by breaking high-level visual questions into structured analytic steps and highlighting semantically relevant chart areas. Leveraging large language and vision-language models, we evaluate the system on 45 tasks across 12 chart types and validate its outputs with trained visualization users, confirming its ability to produce interpretable and expert-aligned reasoning sequences.
}


\teaser{
  \vspace{-20pt}
  \centering
  \includegraphics[width=\linewidth, alt={}]{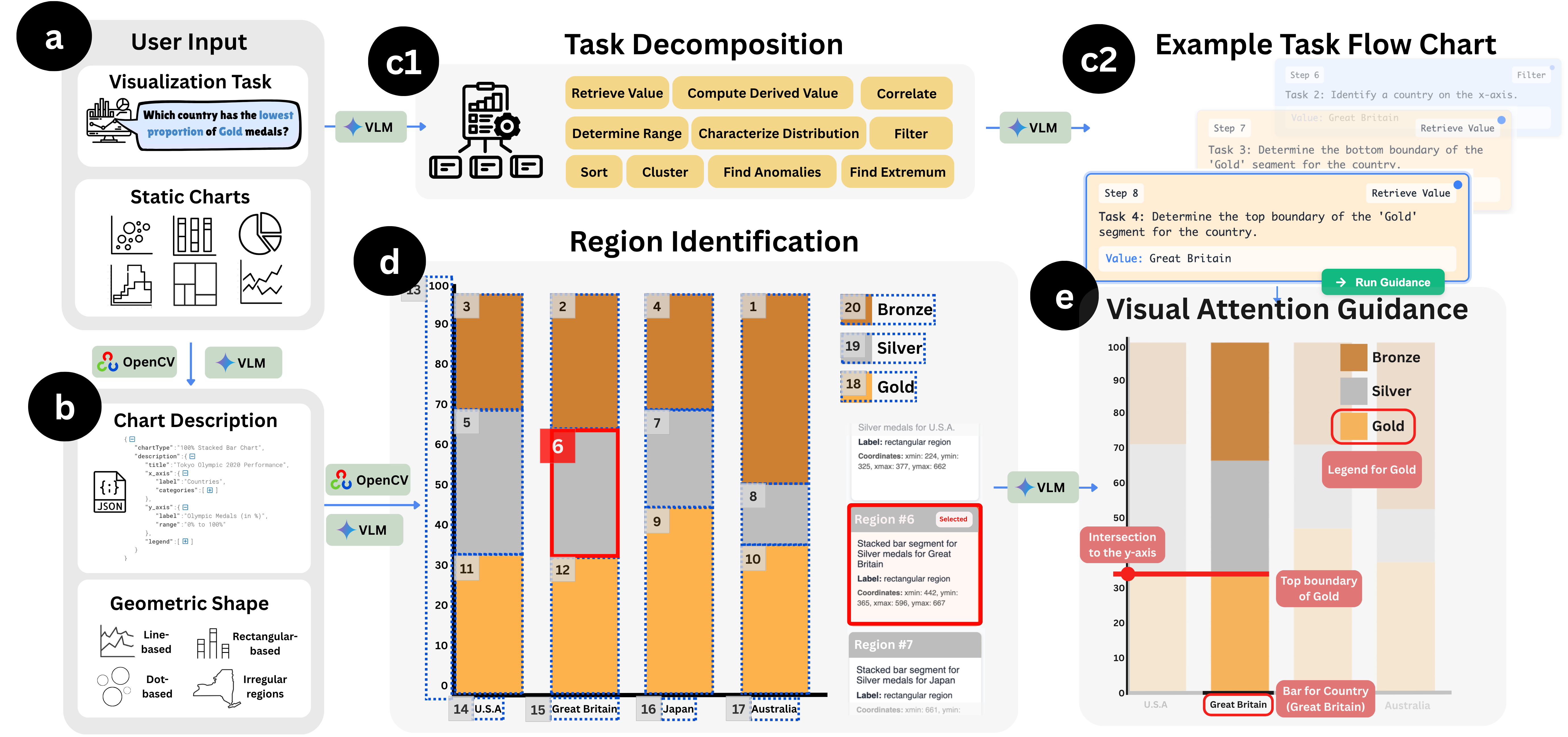}
  \vspace{-20pt}
  \caption{%
  	\textbf{A pipeline overview of ViStruct.} \textit{(a)} User provides a visualization task description and static charts. \textit{(b)} Behind the scenes, the system generates a detailed textual description of the chart and identifies its geometric elements. \textit{(c1)} The high-level task is decomposed into a sequence of low-level component subtasks. \textit{(c2)} An example shows how the original task is broken down into a step-by-step reasoning flow. \textit{(d)} The system detects and labels chart regions based on geometric features and spatial relationships.\textit{(e)}  The chart is automatically annotated with task-relevant Areas of Interest (AOIs) to guide user attention.%
  }
  \label{fig:teaser}
}

\renewcommand{\manuscriptnotetxt}{}

\usepackage{tabu}                      
\usepackage{booktabs}                  
\usepackage{lipsum}                    
\usepackage{mwe}                       

\usepackage{mathptmx}                  

\begin{document}


\firstsection{Introduction}
\maketitle

Decomposing a complex task into smaller, manageable components is widely recognized as an effective strategy for problem-solving~\cite{briakou2024, Kazemitabaar2024Improving, stepbystep, Morrison2015Subgoals}. In many domains, expert users perform such decomposition intuitively, applying structured strategies to reason through problems efficiently. However, these strategies are often implicit and difficult to observe, replicate, or teach~\cite{Nelms2019ExpertNovice}. Consequently, by simulating how experts break down complex tasks, we can externalize these internal reasoning patterns and make them accessible for analysis, instruction, or further automation~\cite{Tullis2015Scheduling, Kazemitabaar2025Improving}. 

Data visualization tasks are a prime example of this need. They involve multiple cognitive stages and high-level goals~\cite{keim2002information}, such as identifying trends or comparing proportions, which are typically achieved through a sequence of low-level perceptual and analytic operations~\cite{amar_low-level_nodate}. Experts, defined as experienced users with demonstrated fluency in interpreting standard chart types, typically perform these reasoning steps fluidly and implicitly. Consequently, their structured reasoning remains hidden, making it challenging to study, explain, or teach~\cite{karer_2020, teo_exploratory_2023}. This highlights a significant gap in capturing and modelling expert reasoning processes; bridging this gap is crucial for developing more interpretable visualization systems and laying the foundations for educational tools.

Despite this need, generating expert-level reasoning for visualization analysis remains challenging.  Task decomposition must be tailored to the structure and semantics of each chart, and each analytic step must be precisely linked to the relevant visual regions. Manually crafting these reasoning is infeasible at scale due to the wide variability in visualization goals, subtasks, and chart designs. Although AI-driven approaches have successfully decomposed complex tasks in other domains \cite{tankelevitch2024metacognitive, suh2023sensecape}, their potential for decomposing visualization-specific tasks and delivering precise visual attention guidance remains largely unexplored.

To bridge this gap, we introduce \textbf{ViStruct}, an automated pipeline that simulates expert-like visual reasoning through structured task decomposition and region-based visual attention.  ViStruct leverages large-language (LLM) and vision-language models (VLM) to systematically \textbf{decompose visualization tasks}, dynamically \textbf{identify context-specific AOIs}, and effectively \textbf{produce actionable instructional sequences} tailored to the structure and semantics of the chart.

To validate the scalability of the proposed technique, we applied it to 45 visualization tasks across 12 chart types. We evaluated our approach with 20 domain experts, confirming the expert-like guidance and transparency provided by ViStruct. The resulting technique is publicly available as an open-source interactive platform for researchers and practitioners, accessible \href{https://vi-struct.vercel.app/}{here}.

\section{Related Work}
Our work draws upon related efforts in AI-driven reasoning frameworks for task decomposition and in vision-language approaches for spatial and semantic understanding of data visualizations. 
\subsection{AI Assisted Task Decomposition}


Previous research has investigated how AI can assist in task decomposition. Techniques like Chain-of-Thought prompting~\cite{wei2022chain} and ReAct~\cite{yao2023react} help structure reasoning through sequential steps and action-based feedback. These methods are often embedded in systems such as Talk2Data~\cite{guo2024talk2data} and LightVA~\cite{zhao2024lightva} to support interactive visual analysis and analytic planning. Additionally, some approaches emphasize user control and trust by letting users refine or verify the decomposition through interactive steps~\cite{Kazemitabaar2024Improving}, promoting transparency and interpretability.

In the context of data visualization, explainable AI frameworks~\cite{yin2024data} and workflow automation tools~\cite{alves2024ai} follow a reactive paradigm, where users must explicitly request task decomposition or guidance. This reactive design places the burden of initiative and strategy on the user, which deviates from how experts naturally guide others through analysis. Rather than waiting for user prompts, ViStruct anticipates the visual analytic process by automatically breaking down the visualization task and highlighting relevant chart regions in a meaningful order, mirroring the reasoning that experts employ.

\subsection{Region-Aware Processing in Data Visualization}
Recent work reveals that while VLMs show some capability in interpreting chart structures and high-level relationships \cite{huang2024relationvlm}, they often lack consistency and robustness \cite{mukhopadhyay2024unraveling}. Many models struggle with understanding visual language and fail to capture relational information accurately, which is critical for interpreting charts \cite{hou2024vision}. Even advanced models still face challenges in reliably extracting meaningful visual relationships.

Newer approaches emphasize region-based understanding. MapReader\cite{zhang2025mapreader} demonstrates how spatial visualizations benefit from region-level segmentation, while intermediate text representations\cite{ye2024beyond} have been used to improve flowchart comprehension. With the help of \hyperlink{https://opencv.org/}{OpenCV}, the Chain-of-Region technique\cite{lichain} proposes explicitly segmenting charts into interpretable regions (axes, bars, legends, etc) and combining this with VLMs enables precise coordinate-to-region mapping, which is crucial for tasks like identifying the correct values in bar or line charts. 

ViStruct leverages region-based techniques by detecting semantically meaningful regions within charts and defining task-specific AOIs. Motivated by sequential visual cues (SVCs) \cite{sequantial}, which guide attention through critical regions in order, ViStruct integrates region segmentation directly into VLMs, aligning visual cues with each step of the task decomposition.

\vspace{-2mm}
\section{Design Goals} \label{goals}

In designing ViStruct, we aimed to simulate expert-like reasoning in visualization tasks, particularly how experts interpret charts step by step and focus on relevant visual regions. To derive these design goals, we extensively reviewed visualization and cognitive science literature and analyzed known distinctions in how experts and novices approach visual reasoning. Prior studies show that experts systematically attend to semantically meaningful regions, interpret visual encodings through structured reasoning sequences, and integrate spatial and semantic cues to guide their analysis~\cite{moerth_kostic_gehlenborg_pfister_beyer_nobre_2025, brunye_eye_2020, nobre_24, Rezaie2024}.
The design goals for ViStruct are:\\
\textbf{G1: Semantic Region Understanding} During the encoding stage of visualization comprehension, experts naturally identify and interpret distinct semantic regions of a chart (axes, data marks, and labels) to extract relevant visual information \cite{brunye_eye_2020}. Effective encoding requires accurately understanding each region's roles within the visual structure \cite{nobre_24}. Therefore, the system must detect these components accurately and assign labels meaningfully reflecting each component's function. \\
\textbf{G2: Precise Coordinate Mapping} In the decoding stage, the visual elements identified must be translated into their corresponding quantitative meanings. This process involves precisely mapping spatial features, such as the height of bars or the positions of data points, to numerical values through alignment with reference regions and axis scales \cite{nobre_24}. The system must support this decoding step so that users can interpret the quantitative information embedded within the spatial arrangement of chart components.\\
\textbf{G3: Structured Task Decomposition} 
Experts often approach visualization tasks by intuitively breaking them down into smaller analytic steps without making this process explicit~\cite{amar_low-level_nodate}. This decomposition happens internally and fluidly, informed by experience and familiarity with visual encoding strategies~\cite{Kerren2020}. ViStruct aims to simulate this expert behaviour by externalizing the reasoning path: it generates a structured sequence of subtasks that mirrors the analytic flow an expert might follow.\\
\textbf{G4: Supporting Diverse Data Encoding}
Charts use a wide range of symbols, layouts, and dimensional mappings to encode data; each may call for a different approach to interpretation\cite{boner2019}. To support diverse chart types and visualization tasks, the system must accurately classify chart components by integrating information from both textual descriptions and geometric shapes.\\
\textbf{G5: Interpretability and Transparency}
Every stage of the reasoning process should be clearly explained to improve interpretability. This includes how regions are divided, what data is extracted, and why certain decisions are made. ViStruct enhances this process using visual attention guidance \cite{jamet_attention_2008, xie_more_2017}, such as highlights and circled AOIs, to direct users toward task-relevant areas. These visual cues allow users to follow the model's interpretation path.

\vspace{-3mm}
\section{ViStruct}
Considering the design requirements, we developed ViStruct as a prototype system to explore automated, expert-like reasoning of visualizations through chart decomposition and visual guidance. This section outlines the ViStruct pipeline and its integration, beginning with a user-facing scenario and then detailing each system component: chart characterization, task decomposition, region annotation, and attention-guided output. For each stage, we describe how it supports the design goals introduced in Section~\ref{goals}.

ViStruct is implemented as a prototype interactive platform using TypeScript and the Next.js framework, with a backend powered by OpenCV for region detection. The system is model-agnostic and adaptable to different vision-language models. In our initial experiments,We tested Gemini-2-Flash\footnote{Gemini-2-Flash: https://deepmind.google/technologies/gemini/flash}, GPT-4V\footnote{GPT-4V: https://openai.com/research/gpt-4v-system-card}, and Claude 3.7\footnote{Claude 3.7: https://www.anthropic.com/news/claude-3-7-sonnet}. on structured chart inputs at each stage of the pipeline. We selected Gemini-2-Flash for integration due to its faster response time and more reliable performance in understanding visual elements.

\vspace{-3mm}
\subsection{Overview and Usage Scenario}

ViStruct simulates how chart-literate experts approach visual reasoning tasks, especially those in visual literacy assessments such as VLAT\footnote{VLAT: https://bckwon.pythonanywhere.com/}. These tasks are intended not to extract novel insights but to train users in structured interpretation strategies that experts implicitly use. ViStruct externalizes these strategies through interactive, step-by-step visual guidance.


Given a static chart and a user-defined visualization question, ViStruct processes the input through multiple stages (Fig.\ref{fig:teaser}). It begins by identifying the chart type and extracting structural elements (e.g., bars, axes, labels) using OpenCV and Gemini-2-Flash (Sec.\ref{sec:chart_characterization}), supporting \textbf{G1: Semantic Region Understanding}. This structured representation is passed to an LLM-driven pipeline to decompose the question into low-level analytic subtasks (Sec.\ref{sec:task_decomposition}), supporting \textbf{G3: Structured Task Decomposition}. Subtasks are organized into a coherent, editable flow (Sec.\ref{sec:decomposition_flow}), ensuring transparency (\textbf{G5}). Concurrently, precise visual regions associated with each subtask are identified and annotated (Sec.~\ref{sec:region_identification}), enabling accurate coordinate mapping (\textbf{G2}). Users progress step-by-step with interactive visual cues highlighting these Areas of Interest (AOIs).

To demonstrate how ViStruct operates in practice, we present a scenario inspired by the VLAT questionnaire, where a user analyzes a chart of Olympic medals (see Fig. ~\ref{fig:teaser}). The chart is a 100\% stacked bar chart that displays each country's distribution of medals: gold, silver, and bronze. The user answers the question: \textit{“Which country has the smallest proportion of gold medals?”} Although this question may seem straightforward, answering it visually involves several steps. The user must identify the gold medal segments, estimate their relative height in proportion to the total height of the bar, and make comparisons across different countries while disregarding other visually present but irrelevant data.

ViStruct automatically parses the chart into semantic regions such as bars, axes, and legends, supporting \textbf{G1}. Users can explore the chart interactively (e.g., hovering over a segment shows: \textit{Gold medal count for USA}). The system decomposes the task into a structured flow (e.g., isolate gold segments, read segment heights, normalize by bar total, compare proportions), surfacing the kind of stepwise reasoning that experts typically perform internally, supporting \textbf{G3}. To assist decoding, the system draws boundary lines and reference overlays (Fig.~\ref{fig:teaser}e), enabling accurate interpretation of visual encodings (\textbf{G2}). Finally, each subtask is editable, and the whole reasoning sequence is transparent, supporting \textbf{G5}.


\vspace{-5pt}
\begin{figure}[htb]
  \centering
  \includegraphics[width=1\linewidth]{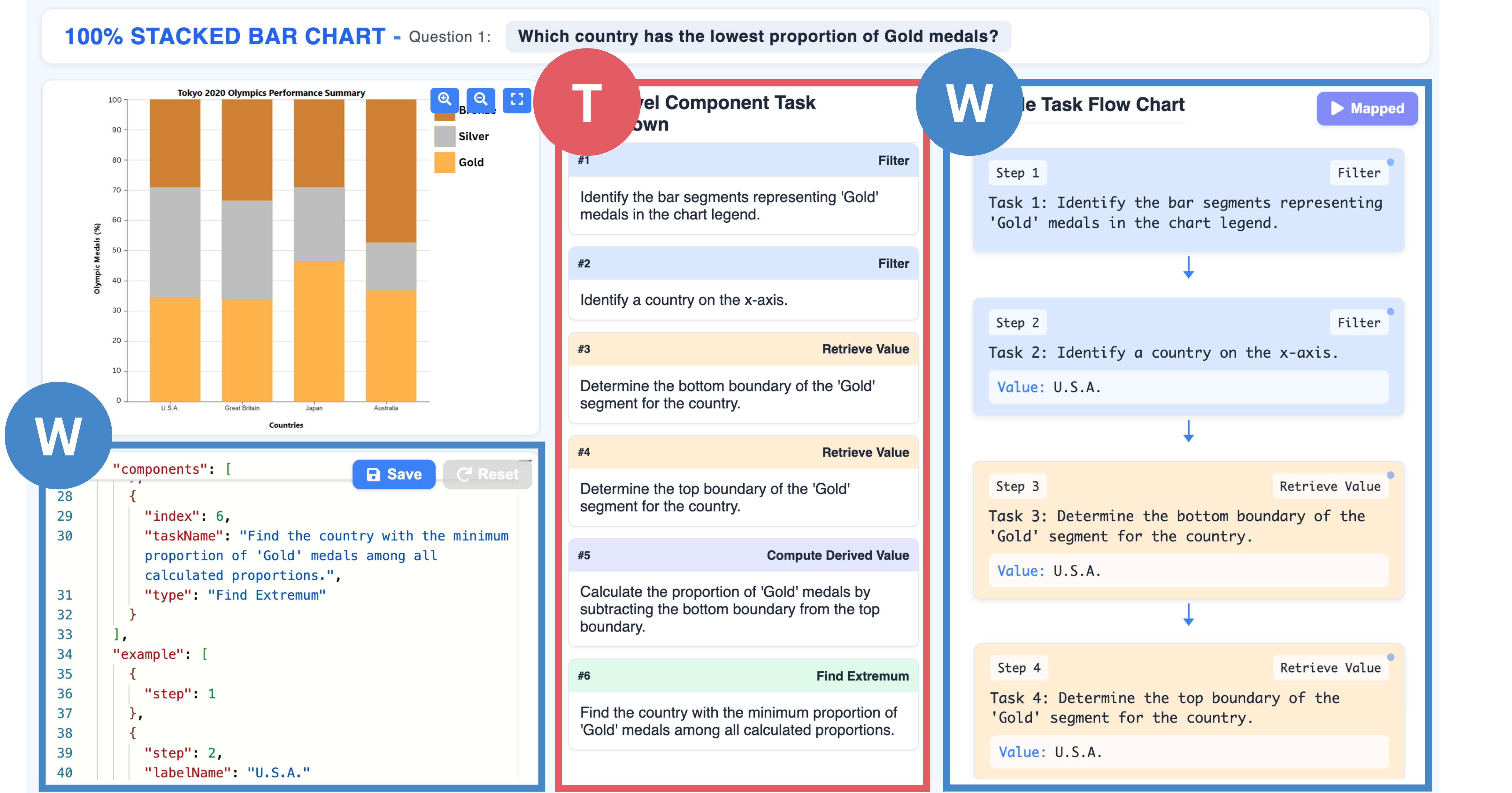}
  \caption{\label{fig:decomp}
           \textbf{(T) Task Decomposition:} The system presents a low-level component task breakdown derived from the user’s high-level question, categorized by task type. \textbf{(W) Editable Workflow Interface:} Users can view and manipulate the task structure and customize the workflow, selecting variable instances and reordering steps.}
\end{figure}

\vspace{-15pt}
\subsection{Chart Characterization} \label{sec:chart_characterization}
ViStruct begins by characterizing the input chart to extract its structural and semantic layout as a foundation for downstream reasoning. Gemini-2-Flash organizes them into a structured JSON representation, including axis ranges, variable names, data groupings, and encoding types.

To improve the reliability of OpenCV-based region detection, graphical elements are categorized into four shape types: line-based, dot-based, rectangular, and irregular. This classification allows ViStruct to adapt region analysis methods to the chart’s visual encoding scheme. Chart characterization supports \textbf{G4} by enabling ViStruct to generalize across diverse chart types and visual conventions.

\subsection{Task Decomposition} \label{sec:task_decomposition}

ViStruct uses a multi-stage prompting framework with three sequential prompts to decompose high-level user tasks into low-level analytic subtasks. The LLM operates on a structured input that includes (1) the user's natural language query, (2) the JSON chart description (Sec.~\ref{sec:chart_characterization}), and (3) a list of labelled chart regions with spatial metadata.

Decomposition is guided by a predefined taxonomy of ten low-level task types\cite{amar_low-level_nodate}. The first prompt produces a breakdown based on this taxonomy. The second refines each step to ensure it is grounded in specific, executable, and sufficiently detailed chart regions. If a step is too abstract, the LLM splits it into atomic components. The third prompt validates the structure, resolves step dependencies, and removes redundant subtasks (e.g., repeating the same operation across different chart elements).

This breakdown–refine–verify process ensures each step is precise, executable, and aligned with the task structure. It directly supports \textbf{G3} by enabling structured task decomposition and \textbf{G4} by adapting reasoning to diverse chart types and visual encodings.

\vspace{-3pt}
\subsection{Decomposition Flow Example} \label{sec:decomposition_flow}
Due to visual analysis's compositional nature, there is often no correct way to sequence low-level operations. ViStruct presents one example workflow based on the decomposition results (Fig.~\ref{fig:decomp}), illustrating a valid approach. Recognizing that users may prefer different reasoning paths, the system allows them to modify the sequence, adjust parameters, or reorganize tasks to fit their analysis strategy better. The LLM then validates the revised flow to ensure logical consistency. If valid, both the workflow and the underlying decomposed tasks are updated accordingly.

\begin{figure}[htb]
  \centering
  \includegraphics[width=1\linewidth]{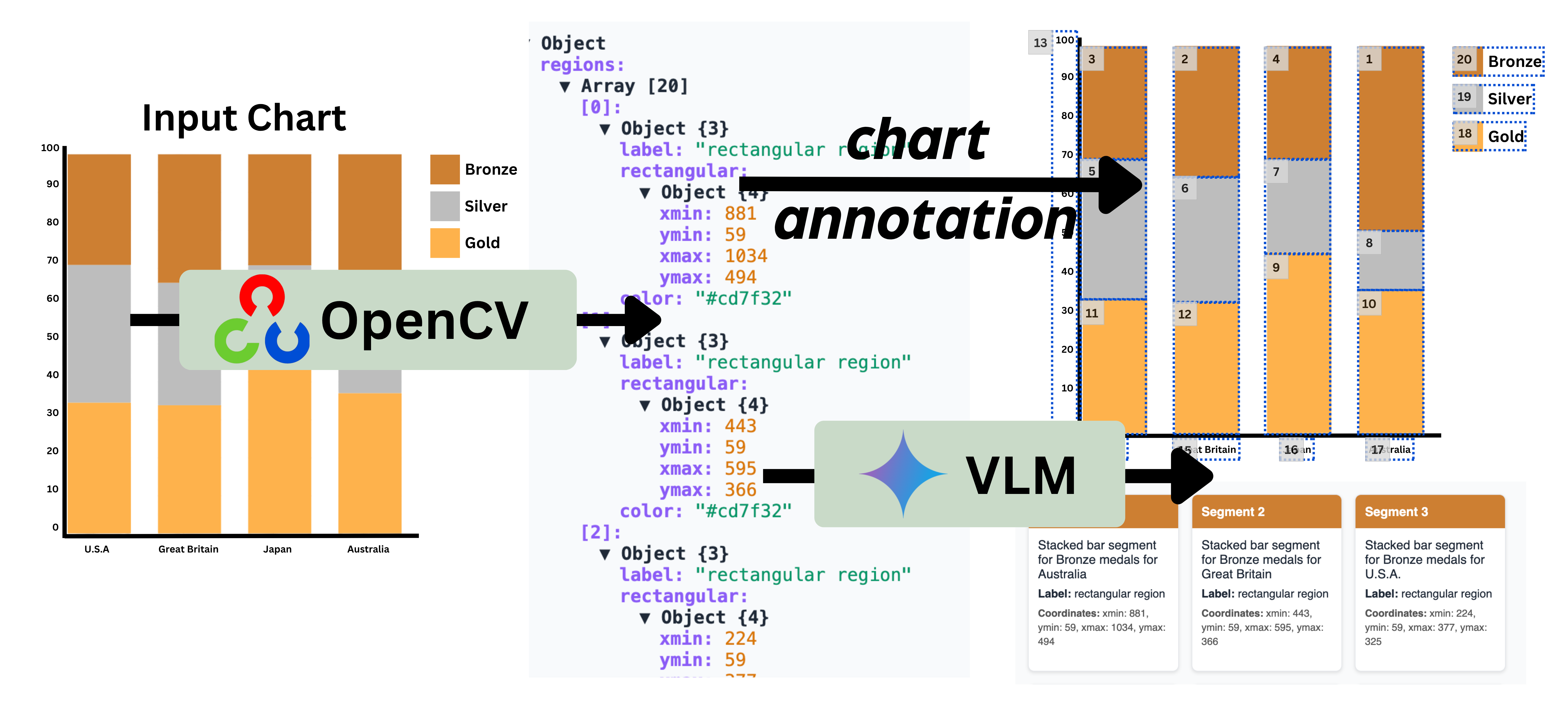}
  \caption{\label{fig:region}
    \textbf{Workflow of region identification}: The input chart is processed using OpenCV to identify distinct regions corresponding to chart segments. These regions are labeled by the system. VLM then identifies each annotated region with meaningful descriptions.}
\end{figure}

\vspace{-6mm}
\subsection{Region Identification} \label{sec:region_identification}
Region identification is a key step in the ViStruct pipeline, providing the precise coordinate-level data needed for visual guidance and accurate annotation. To highlight AOIs during task execution, the system must first detect all meaningful chart components, including axes, labels, data marks, and legends.

The process begins with OpenCV-based text detection (Fig.~\ref{fig:region}), which identifies individual characters and outputs their bounding boxes (\texttt{x-min}, \texttt{x-max}, \texttt{y-min}, \texttt{y-max}). Characters with similar x- or y-values are grouped into complete text elements, allowing the system to reconstruct axis labels, ticks, and legend entries.

For non-text elements, OpenCV’s edge detection and colour segmentation identify visual regions such as bars, pie slices, or legend swatches based on shape boundaries and colour differences. In charts with continuous marks like lines or areas, the system uses identified axis ticks as reference points and interpolates pixel coordinates along the visual path to map data values.

This yields a complete set of region coordinates but without semantic meaning. To bridge this gap, the system overlays numbered labels on an annotated chart image and sends coordinate data to Gemini-2-Flash, which returns natural language descriptions for each region (e.g., \textit{"Gold medal for the USA"}). This transforms low-level visual segments into semantically meaningful components, directly supporting \textbf{G1} by identifying functional chart regions and \textbf{G2} by linking visual geometry to interpretable data values.

\vspace{-2mm}
\subsection{Visual Attention Guidance} \label{sec:visual_attention}
After chart regions are semantically identified, ViStruct generates step-by-step visual guidance to help users focus on relevant elements and understand how to extract the correct information from the chart.

To generate this guidance, the system provides the LLM with structured input: (1) the current low-level subtask, (2) a JSON object detailing chart regions, their coordinates, region IDS, and semantic labels, and (3) task-specific metadata from stage \ref {sec:chart_characterization}. Each of the subtask types uses an individual prompt to ensure contextual accuracy and clarity.

Based on this input, the LLM generates a sequence of visual guidance steps tied to specific regions or reference points. Some, like bar segments or axis labels, are directly retrieved from the JSON, while others involve simple geometric reasoning. For instance, the system draws a horizontal line from the bar’s edge to the y-axis to read a bar's top value, guiding the user’s attention.

ViStruct creates an interpretable and grounded guidance flow by combining textual instructions with precise visual markers. This directly supports \textbf{G2} by helping users connect spatial features to quantitative meaning and \textbf{G5} by making each reasoning step transparent and visually traceable.

\vspace{-2mm}
\section{Evaluating the ViStruct Pipeline}
We conducted two evaluations to assess ViStruct’s effectiveness as an expert-like visual reasoning pipeline: a performance evaluation to test decomposition accuracy and scalability and an expert evaluation to understand user perceptions regarding clarity, usefulness, and expert-like behaviour.

\vspace{-2mm}
\subsection{Performance Evaluation} 
\vspace{-1mm}
We tested ViStruct on 45 visualization tasks drawn from \hyperlink{https://bckwon.pythonanywhere.com/}{VLAT} and \hyperlink{https://tools.visualdata.wustl.edu/experiment/\#/}{Mini-VLAT}, covering all 12 chart types (\textbf{G4}). Each task was executed five times, resulting in 225 trials. For each trial, we evaluated ViStruct’s ability to identify semantic chart regions (\textbf{G1}), map visual features to data values (\textbf{G2}), and generate coherent task decompositions (\textbf{G3}). ViStruct produced correct outputs in 192 of the 225 trials (85.33\%). It performed consistently on concrete tasks such as value lookup and filtering. In contrast, more abstract tasks such as correlation analysis in multidimensional charts (i.e., bubble charts) occasionally showed inconsistencies by selecting the wrong visual channels for analysis.

\vspace{-2mm}
\subsection{Expert Review}
\vspace{-1mm}
 We conducted an expert evaluation to assess whether ViStruct demonstrates expert-like behaviour and provides valuable guidance. Each participant was assigned three charts and could select any task from the question bank. They evaluated ViStruct's decomposition, workflow logic, region identification, and visual guidance. Specifically, they rated ViStruct on: \textit{(1)} usefulness for novice users, \textit{(2)} accuracy of subtasks and AOIs, and \textit{(3)} alignment with their reasoning. Participants also gave open-ended feedback and used a think-aloud protocol during tasks.

 We recruited 20 participants (M=12, F=8), including 12 undergraduates who completed a data visualization course, 4 graduate researchers, and 4 industry analysts. All were screened for chart familiarity and reported an average expertise rating of \textit{6.35/7}. Each session lasted 30 minutes, and participants received a \$10 gift card.

\textbf{Experts' Perceptions and Feedback.}
Participants rated ViStruct highly for guiding visual reasoning (\textit{M = 6.14, SD = 1.38}); the accuracy of its decompositions and AOIs scored 5.93 (\textit{SD = 1.58}), and its perceived expert‑likeness was 5.97 (\textit{SD = 1.53}).

\textbf{Step-by-step guidance supported participants in organizing their reasoning.} Several users noted that the combination of visual overlays and sequential explanations clarified how to approach complex charts. One participant remarked that “\textit{visual overlays alongside explanatory text are a useful step‑by‑step guide... especially with bubble charts where there are more variables than people are used to}” (\textit{p14}). This form of guided interaction helped participants build a mental model of the task structure (\textbf{G2, G5}).

\textbf{Region annotations anchored users' attention to relevant chart elements.} Participants appreciated the system’s ability to highlight and label specific visual components (\textbf{G1}). Numbered regions were especially helpful for less experienced users, who found the annotations reduced confusion (\textbf{G5}). One participant commented that “\textit{the mapping feature... helps visually guide them in each step by highlighting the regions to focus on}” (\textit{p9}).

\textbf{Experts interpreted the decomposition as instructional guidance aimed at novice users.} While many agreed that the step-by-step breakdown resembled how they would teach a novice (\textbf{G3}), they did not feel the need to follow every step themselves. One participant observed that “\textit{the steps were clear and resemble how a domain expert would interpret and guide someone}” (\textit{p3}). In contrast, some others noted that “\textit{for experienced people, some of the steps might be a bit too detailed, but someone who is new to these charts could find it very helpful}” (\textit{p8}).


\vspace{-1mm}
\section{Limitations and Future Work}
\vspace{-1mm}
\textbf{Varying Effectiveness of Guidance Across Task Types:} Our evaluation showed that the AOI-based guidance works well for concrete tasks such as filtering or locating values, where visual cues are direct. However, AOIs alone are less effective for more abstract tasks (i.e., correlation). These tasks require integrating multiple elements, suggesting that richer cues, such as tooltips or side panels, may be more appropriate. Future work should explore customized guidance strategies through participatory design or user feedback.
\\\textbf{Supporting Diverse Reasoning Paths and Scaffolding:} While the ViStruct decomposition strategy is effective for many tasks, it may not always match how users reason about complex visualizations. Abstract tasks often permit multiple valid approaches, so a fixed breakdown can be limiting. Future work should enable flexible, user‑driven reasoning by adding an interactive chatbot that refines task flows, clarifies goals, and adapts strategies to each chart. In addition, the system could benefit from providing contextual explanations that clarify the rationale behind each step, helping users understand not just what to do, but why certain visual reasoning strategies are effective.
\\\textbf{Human-in-the-loop Opportunities and Alignment with Novice Intentions:}  A significant portion of ViStruct’s failures in visual attention guidance stemmed from errors in chart region identification, particularly due to limitations in OpenCV-based detection (e.g., missed or fragmented visual elements). These failures impact the quality of AOI mapping and, by extension, the effectiveness of task decomposition. Rather than relying solely on automation, we see an opportunity to introduce a human-in-the-loop setting where interactive correction of detected regions could reduce such errors while simultaneously supporting learning. This form of productive friction~\cite{Kazemitabaar2025Improving} may help novices develop a deeper understanding of visualization literacy by engaging directly with the structural interpretation of charts.

\vspace{-2mm}
\section{Conclusion}
\vspace{-1mm}
In conclusion, we present \textbf{ViStruct}, an automated pipeline that simplifies visualization tasks by decomposing them into semantically meaningful subtasks, extracting structured region information, and providing step-by-step visual guidance. The prototyped system is deployed and can be accessed \href{https://vi-struct.vercel.app/}{here}. Our evaluation shows reliable performance across diverse static charts and tasks. We plan to enhance flexibility with context-aware guidance and extend the approach to interactive settings to better support insight extraction.

\bibliographystyle{abbrv-doi}

\newpage

\acknowledgments{
\vspace{-5pt}
This work was conducted in accordance with Research Ethics Board (REB) Protocol \#48124. We sincerely thank Runlong Ye and Matthew Verona for their thoughtful input and valuable feedback throughout the development of this work.}

\end{document}